\documentclass[aps,prl,twocolumn,superscriptaddress,longbibliography]{revtex4-1}

\usepackage{graphicx}
\usepackage{bm}
\usepackage{hyperref}
\usepackage{todonotes}
\usepackage{color}
\usepackage{comment}
\usepackage{verbatim}
\usepackage{soul}
\usepackage{braket}
\usepackage[acronym]{glossaries}

\def\BAIO{Ba$_5$AlIr$_2$O$_{11}$}

\newacronym{RIXS}{RIXS}{resonant inelastic x-ray scattering}
\newacronym{ARPES}{ARPES}{angle-resolved photoemission spectroscopy}
\newacronym{2D}{2D}{two-dimensional}
\newacronym{SOC}{SOC}{spin-orbit coupling}
\newacronym{DFT}{DFT}{density functional theory}
\newacronym{TB}{TB}{tight-binding}
\newacronym{VASP}{VASP}{Vienna Ab initio Simulation Package}
\newacronym{DPS}{DPS}{Direct Product State}
\newacronym{FWHM}{FWHM}{full width at half maximum}
\newacronym{CF}{CF}{crystal field}

\begin{document}

\title{Direct detection of dimer orbitals in \BAIO{}}

\author{Y. Wang}
\email{yilinwang@bnl.gov}
\affiliation{Department of Condensed Matter Physics and Materials Science, Brookhaven National Laboratory, Upton, New York 11973, USA}

\author{Ruitang Wang}
\affiliation{School of Physical Science and Technology, ShanghaiTech University, Shanghai 201210, China}
\affiliation{Beijing National Laboratory for Condensed Matter Physics and Institute of Physics,
Chinese Academy of Sciences, Beijing 100190, China}

\author{Jungho Kim}
\author{M. H. Upton}
\author{D. Casa}
\author{T. Gog}
\affiliation{Advanced Photon Source, Argonne National Laboratory, Argonne, Illinois 60439, USA}

\author{G. Cao}
\affiliation{Department of Physics, University of Colorado Boulder, Boulder, Colorado 80309, USA}

\author{G. Kotliar}
\affiliation{Department of Condensed Matter Physics and Materials Science, Brookhaven National Laboratory, Upton, New York 11973, USA}
\affiliation{Physics and Astronomy Department, Rutgers University, Piscataway, NJ 08854, USA}

\author{M. P. M. Dean}
\email{mdean@bnl.gov}
\affiliation{Department of Condensed Matter Physics and Materials Science, Brookhaven National Laboratory, Upton, New York 11973, USA}
\author{X. Liu}
\email{liuxr@shanghaitech.edu.cn} 
\affiliation{School of Physical Science and Technology, ShanghaiTech University, Shanghai 201210, China}

\date{\today}

\begin{abstract}
The electronic states of many Mott insulators, including iridates, are often conceptualized in terms of localized atomic states such as the famous ``$J_\text{eff}=1/2$ state''. Although, orbital hybridization can strongly modify such states and dramatically change the electronic properties of materials, probing this process is highly challenging.  In this work, we directly detect and quantify the formation of dimer orbitals in an iridate material \BAIO{} using \gls{RIXS}. Sharp peaks corresponding to the excitations of dimer orbitals are observed and analyzed by a combination of \gls{DFT} calculations and theoretical  simulations based on a Ir-Ir cluster model. Such partially delocalized dimer states lead to a re-definition of the angular momentum of the electrons and changes in the magnetic and electronic behaviors of the material. We use this to explain the reduction of the observed magnetic moment with respect to prediction based on atomic states.  This study opens new directions to study dimerization in a large family of materials including solids, heterostructures, molecules and transient states.

\end{abstract}
\maketitle

Many of the most interesting phases in correlated quantum materials occur in systems with strong Coulomb repulsion $U$, which tends to drive electron localization via the Mott insulating mechanism ~\cite{Keimer2017physics}. Due to this, we often conceptualize the electronic and magnetic properties of these systems in terms of localized states~\cite{Khomskii2016Transition}, even though many of the most interesting cases occur when there is strong competition between $U$ and electron hopping $t$. Of particular interest in this regard is the localized ``$J_\text{eff}=1/2$ state'' in the iridates, which is the conceptual building block for a host of fascinating proposed and observed states including frustrated magnets~\cite{Chaloupka2010Kitaev, Okamoto2007Spin, Balents2010spin, LawlerGapless2008, Zhou2008}, topological insulators~\cite{Pesin2010mott, Shitade2009quantum}, and possibly even unconventional superconductors~\cite{Wang_prl_2011}. Hopping between IrO$_6$ octahedra delocalizes the electrons and is expected to be relevant in many classes of iridate (or other heavy $d$-electron materials) with edge-sharing or face-sharing octahedra ~\cite{Cao2018challenge, panda2015,Ba3InIr2O9_2017, gardner2010_pyrochlore, Streltsov10491_pnas, Ye2018Covalency}. This can heavily modify,  or even destroy, the $J_\text{eff}=1/2$ state motivating arguments about how best to conceptualize the electronic state of iridates~\cite{Mazin2012, IgarashiAnalysis2014, Cao2018challenge}. Directly probing these states is therefore highly desirable. In the simplest case of dimerization between neighboring pairs of IrO$_6$ octahedra one expects the formation of quasi-localized dimer orbitals, which are difficult to probe by photo-emission due to the absence of dispersive bands and hard to probe optically as dipole optical selection rules means that transitions within a orbital manifold are nominally forbidden. \gls{RIXS}, on the other hand, has been shown as an particularly incisive probe of on-site localized transitions in the iridates, but as far as we are aware, has never definitively isolated an excitation associated with dimerization \cite{Liu2012testing, Gretarsson2013Crystal}.

In this Letter, we establish that \gls{RIXS} can directly measure peaks associated with orbital dimerization and that a quantitative description of the dimer electronic configuration can be extracted using \gls{DFT} calculations and multiplet modeling. For this study we employ face-sharing iridiate \BAIO{} and account for how  these interactions modify $J_\text{eff}=1/2$ state expected on the nominally $5d^5$ atom in this crystal, reproducing the previously measured reduction in the magnetic moment \cite{Terzic2015coexisting}. We argue that, given the proven ability of \gls{RIXS} to measure molecules, oxide heterostructures and even ultra-fast transient states, this has great potential to probe orbital dimerization under many different circumstances  \cite{ament2011resonant, Fohlisch2000How, dean2012spin, Lupascu2014Tuning, Dean2016Ultrafast, Lupascu2014Tuning, Meyers2017magnetism}.

\begin{figure}[!ht]
\includegraphics[width=0.45\textwidth]{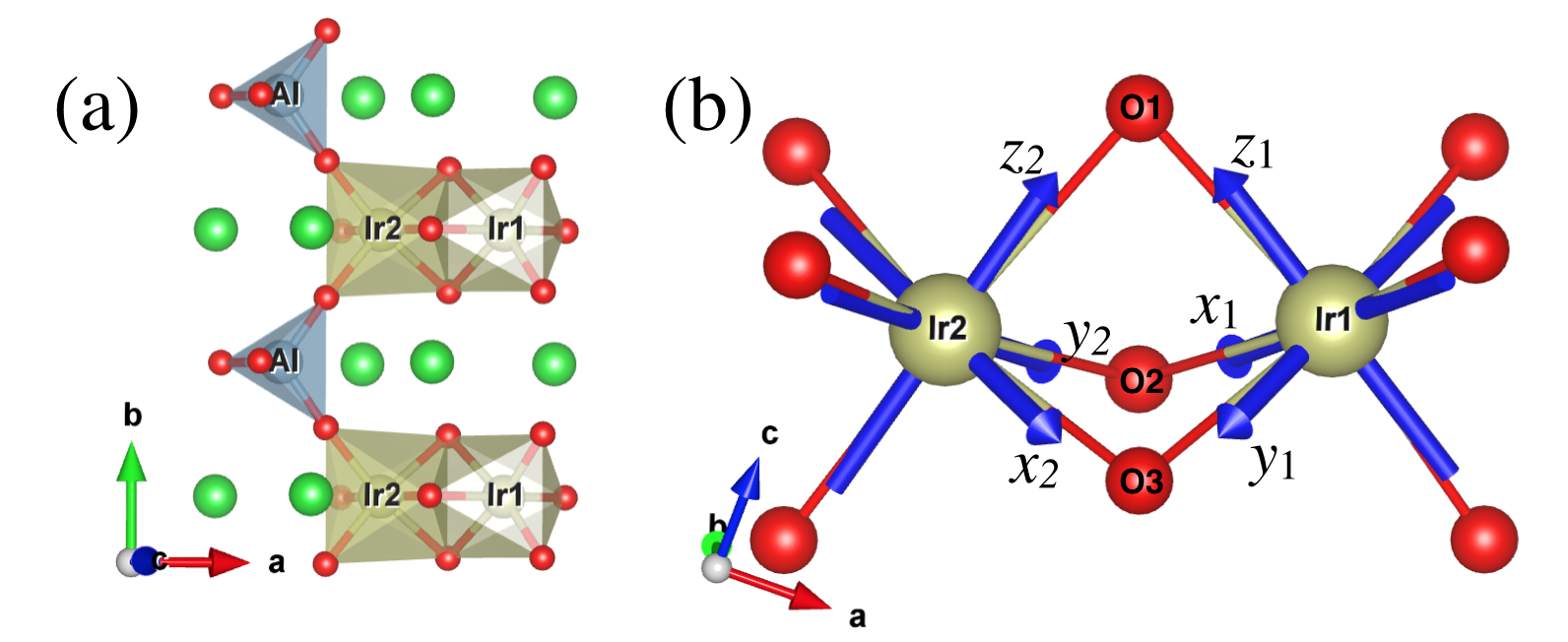}
\includegraphics[width=0.47\textwidth]{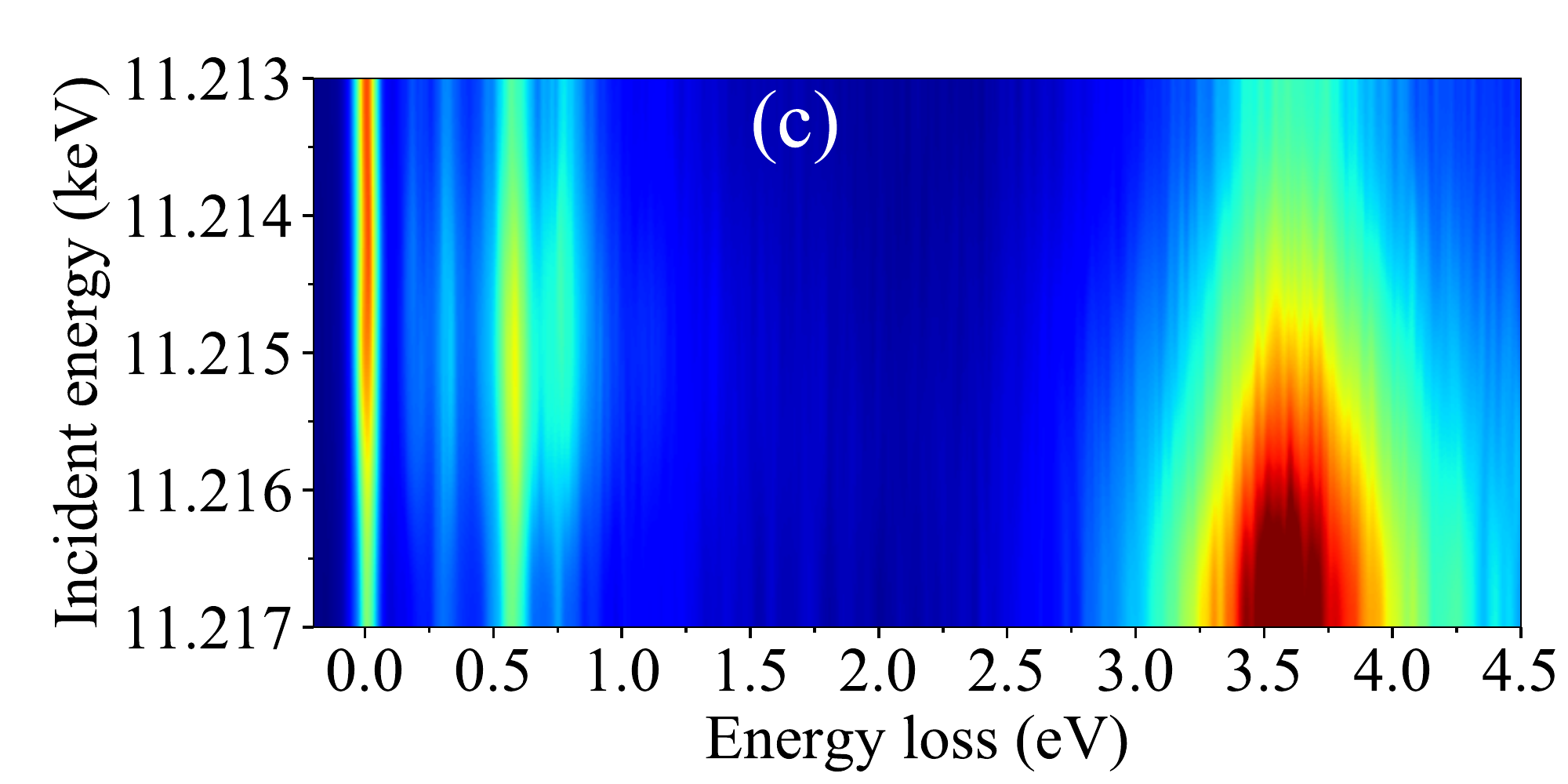}
\caption{(a) Crystal structure of \BAIO{}. The two inequivalent IrO$_6$ octahedra are labeled Ir1 and Ir2 and colored in dark yellow. Quasi one dimensional Ir chain structures occur via the connection of these octahedra through AlO$_4$ tetrahera along the $b$-axis. (b) Illustration of the local Cartesian coordinates used in the definition of $t_{2g}$ orbitals. The orientation of the local coordinates is chosen such that the $x,y,z$-axes are all at an equal angle [$\arccos (1/\sqrt{3})$~rad] with respect to the Ir1-Ir2 bond and the $z$-axis lies in the Ir1-O1-Ir2 plane. The $x,y,z$-axes lie approximately along the Ir-O bonds in this setting.(c) \gls{RIXS} map showing the intra-$t_{2g}$ and $t_{2g} \rightarrow e_g$ excitations.}
\label{pic:struct}
\end{figure}

Single crystal samples of \BAIO{} were synthesized using the self-flux method as described in Refs.~\cite{Terzic2015coexisting,gang2013frontiers}. Previous diffraction and transport measurements confirm high sample quality \cite{Terzic2015coexisting}. \BAIO{} forms an orthorhombic unit cell with space group $Pnma$ (No.~62) with $a=18.76$, $b = 5.755$, and $c = 11.06$~\AA{}~\cite{Terzic2015coexisting}. As shown in Fig.~\ref{pic:struct}(a), two face-sharing IrO$_6$ octahedra form isolated dimers, which are then connected by AlO$_4$ tetrahedra. These two inequivalently coordinated face-sharing Ir octahedra in a dimer are labeled as Ir1 and Ir2, as shown in Fig.~\ref{pic:struct}(a,b). \gls{RIXS} experiments were performed at the Ir-$L_3$ edge by using the MERIX instrument at Sector 27 of the Advanced Photon Source \cite{shvyd2013merix}. A 2~m Si $(8 8 4)$ diced analyzer was used to energy-resolve the scattered x-rays. Different incident beam monochromation setups were tested before settling with a combined $\sim 80$ meV total resolution (Full-Width-Half-Maximum). Data were collected with a horizontal scattering geometry with the incident x-ray polarization parallel to the scattering plane ($\pi$ channel).  The sample was mounted such that $[0 1 0]$  Ir chain direction and the $[1 0 1]$ sample surface-normal are in the scattering plane. Data were collected using an incident x-ray angle $\alpha$ close to $26^{\circ}$ and a detector angle $2\theta$ close to $90^{\circ}$ unless otherwise specified. 

\begin{figure*}[!ht]
\includegraphics[width=0.45\textwidth]{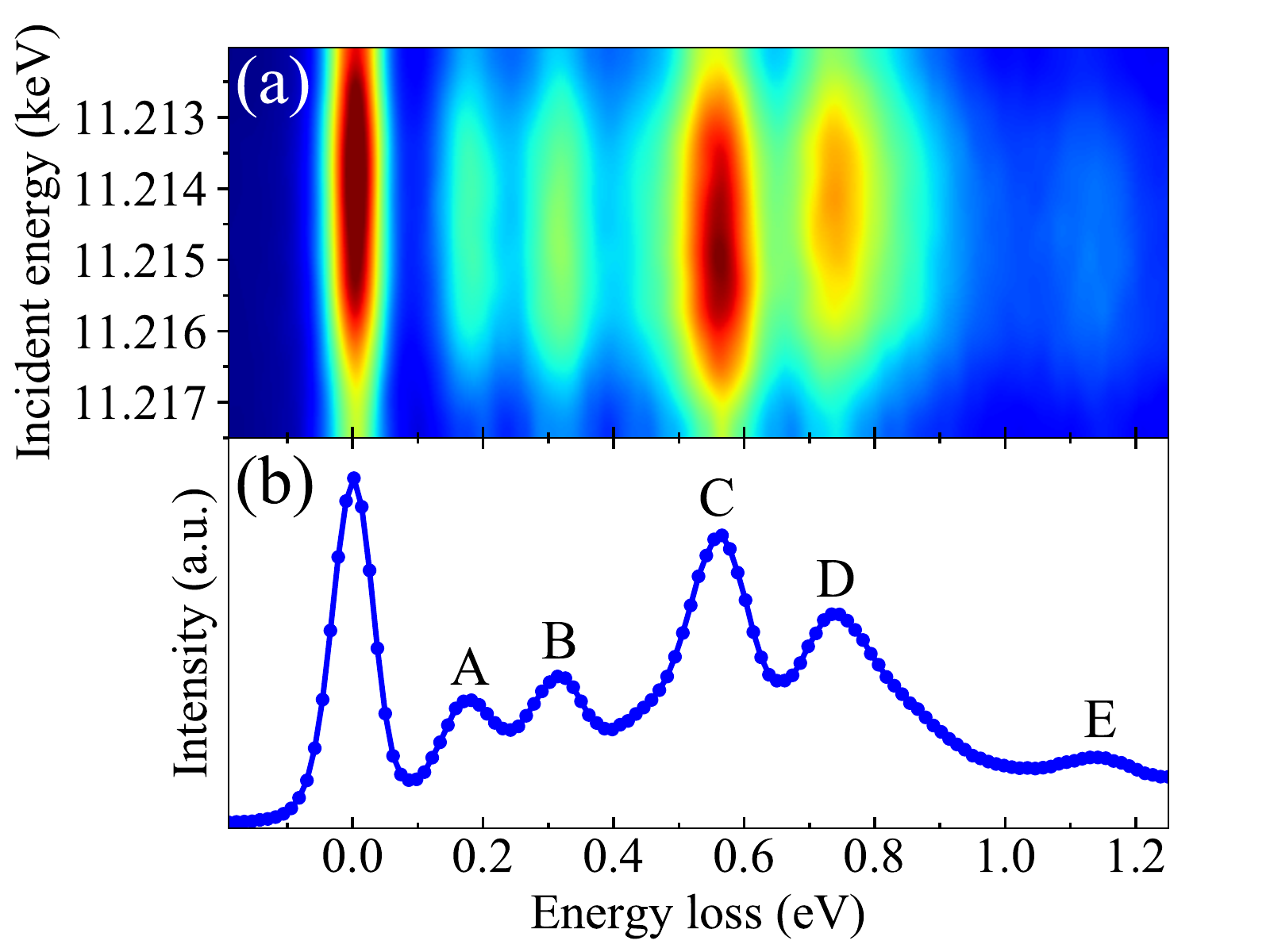}
\includegraphics[width=0.45\textwidth]{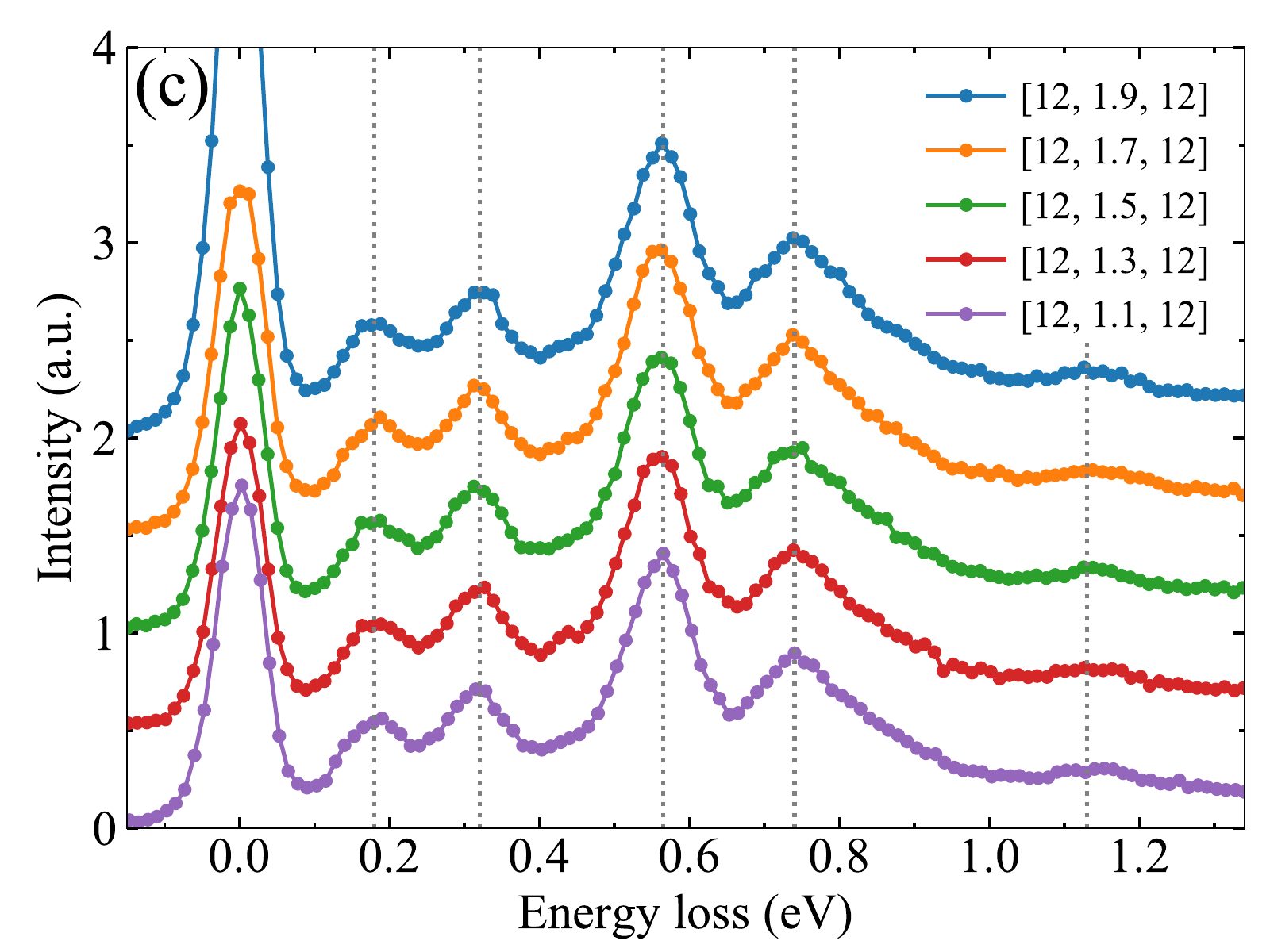}
\caption{(a) High-resolution \gls{RIXS} map at 40~K. (b) The excitation spectrum integrated over $\pm 2$~eV with respect to the resonant energy of 11215~eV. (c) Orbital excitations of \BAIO{} show no significant $\bm{Q}$-dependence indicative of localized states.}
\label{pic:exp_rixs}
\end{figure*}

We initially surveyed incident-energy-dependent \gls{RIXS} at room temperature as shown in Fig.~\ref{pic:struct}(c). The low energy excitations below 1.2~eV resonate at 11.215~keV and can be assigned to intra $t_{2g}$ transitions by comparison to previous work; $t_{2g} \rightarrow e_g$ transitions occur around 3.6~eV and resonant at higher incident energies \cite{Ishii2011Momentum, Liu2012testing, Sala2014Spin}. A more detailed incident energy dependence was mapped at 40~K with 80~meV energy resolution \footnote{Higher resolution data was taken with 30~meV energy resolution, but the this does not reveal any new features.}, focusing on the low energy intra $t_{2g}$ excitations. The spectrum is plotted in Fig.~\ref{pic:exp_rixs}(a,b), where 5 distinct peaks can be clearly identified. They are labeled A-E with energies 0.18~eV, 0.32~eV, 0.565~eV, 0.74~eV and 1.13~eV, respectively. These features show no appreciable dispersive behavior in their peak energies [see Fig.~\ref{pic:exp_rixs}(c)] at various momentum transfer $\bm{Q}$ over almost one Brillouin zone in the chain direction, suggesting that the excitations are confined within a dimer and that coupling through the AlO$_4$ tetrahedra can be neglected on these energy scales. We note that pure spin flip excitations are not observed here, almost certainly because they occur on a too low energy scale.

We use a two-site cluster model for Ir1 and Ir2 [see Fig.~\ref{pic:struct}(b)] to simulate the measured \gls{RIXS} spectrum. The $t_{2g}$ orbitals of Ir1 and Ir2 are defined with respect to the local Cartesian coordinates shown in Fig.~\ref{pic:struct}(b). The ED-RIXS toolkit developed in the COMSCOPE project~\cite{comscope} is used to diagonalize the Hamiltonian and simulate the \gls{RIXS} spectrum, with the real experimental geometry and polarization applied. The energy resolution is set to be 80~meV \gls{FWHM}. More details of the \gls{RIXS} simulation can be found in the Supplemental Materials~\cite{supp}.

\begin{figure*}[!ht]
\includegraphics[width=0.45\textwidth]{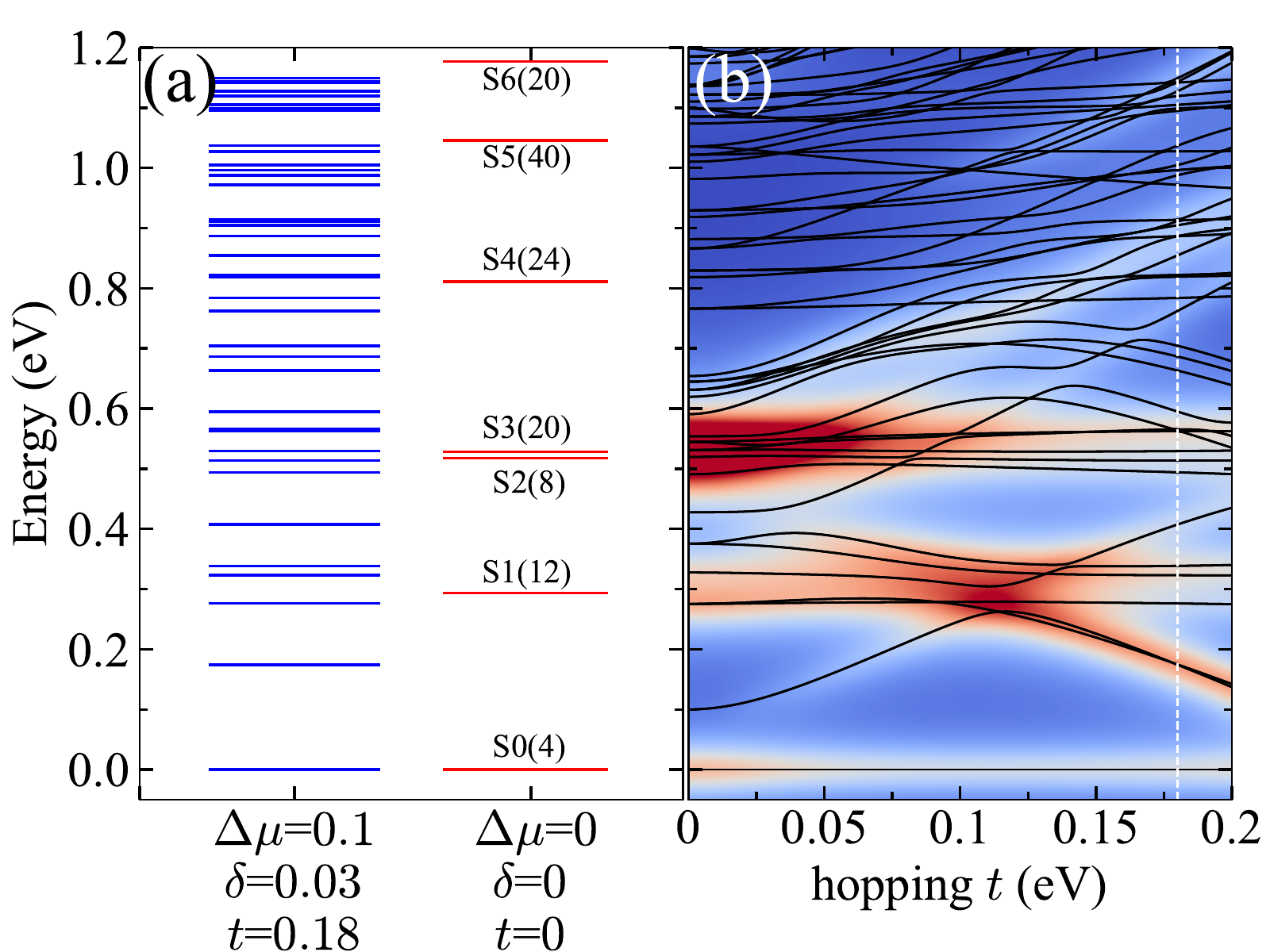}
\includegraphics[width=0.45\textwidth]{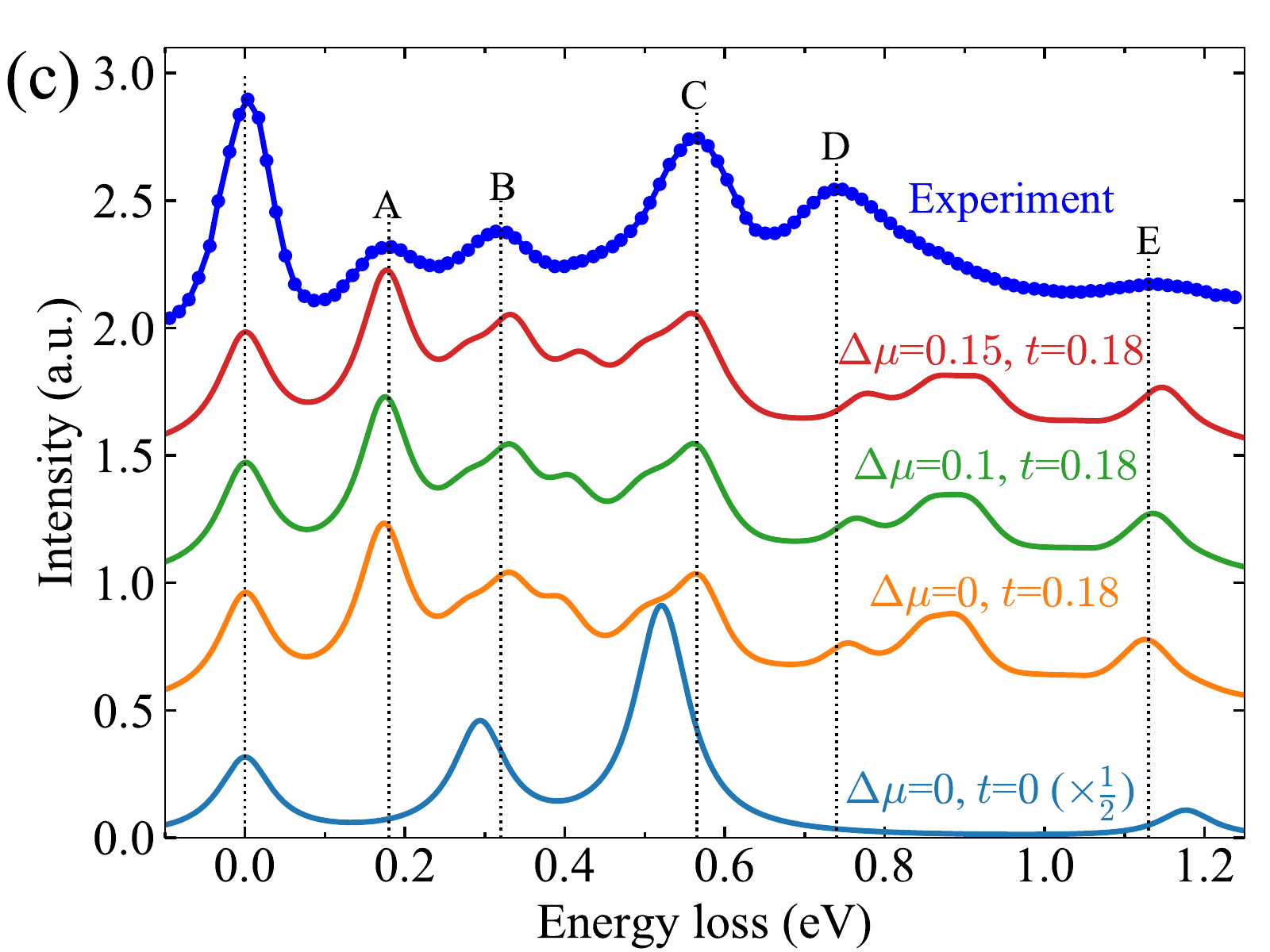}
\caption{(a) Illustration of the splitting of the trivial \acrfull{DPS} after turning on hopping $t$. The letter S is used to label these \gls{DPS} and the numbers in parentheses are their degeneracy. (b) The solid lines are the energy spectrum of the Ir-Ir cluster model as a function of hopping $t$ at $\Delta\mu=0.1$~eV, $\delta=0.03$~eV, and the colormap indicates the corresponding simulated \gls{RIXS} intensity. The vertical dotted line indicates $t=0.18$ eV. (c) The simulated \gls{RIXS} spectrum at several chosen parameters and are compared with the experimental result, $\delta=0$ for the bottom light blue curve and $\delta=0.03$ for all others.}
\label{pic:theory}
\end{figure*}

To understand the electronic structure which gives rise to the observed excitations, we start with the non-physical condition where the Ir1-Ir2 hopping and the non-cubic \gls{CF} are set to zero. With the octahedra decoupled, the electronic structure is determined by the \gls{SOC} $\lambda$, the on-site Coulomb $U$ and the Hund's $J_H$ interactions. We use a $t_{2g}$ Kanamori type Hamiltonian to treat $U$ and $J_H$, which can be written as,
\begin{eqnarray}
\hat{H}_{U}&=&U\sum_{\alpha}\hat{n}_{\alpha\uparrow}\hat{n}_{\alpha\downarrow} +U^{\prime}\sum_{\alpha\neq\beta}\hat{n}_{\alpha\uparrow}\hat{n}_{\beta\downarrow}+U^{\prime\prime}\sum_{\alpha<\beta,\sigma}\hat{n}_{\alpha\sigma}\hat{n}_{\beta\sigma}\nonumber \\
&-&J_{H}\sum_{\alpha\neq\beta}\hat{d}_{\alpha\uparrow}^{\dagger}\hat{d}_{\alpha\downarrow}\hat{d}_{\beta\downarrow}^{\dagger}\hat{d}_{\beta\uparrow}+J_{H}\sum_{\alpha\neq\beta}\hat{d}_{\alpha\uparrow}^{\dagger}\hat{d}_{\alpha\downarrow}^{\dagger}\hat{d}_{\beta\downarrow}\hat{d}_{\beta\uparrow},
\end{eqnarray}
where, $\alpha(\beta)=d_{zx}, d_{zy}, d_{xy}$ is the $t_{2g}$ orbital index.  $\sigma=\uparrow,\downarrow$ is the spin index. $U^{\prime}=U-2J_{H}$ and $U^{\prime\prime}=U-3J_{H}$. With hopping forbidden, the electrons are localized to form $\Ket{d^4}$ on Ir1 and $\Ket{d^5}$ on Ir2, with atomic states $\Ket{d^4, J_{\text{eff}}=0, 1, 2, 2^{\prime}}$ and $\Ket{d^5, J_{\text{eff}}=1/2, 3/2}$, respectively. Excitations can happen within each atomic state sets, and their energies are determined by Hund's coupling and \gls{SOC}. With $J_{H}=0.3$~eV and $\lambda=0.345$~eV, which are consistent with earlier work~\cite{mazin2013,Yuan2017determination,kim2017resonant,paramekanti2018}, the simulated \gls{RIXS} spectrum is shown as the light blue curve in Fig.~\ref{pic:theory}(c) where three peaks are found. The peak near B comes from the excitations from $\Ket{d^{4},J_{\text{eff}}=0}$ to $\Ket{d^{4},J_{\text{eff}}=1}$ states. The peak near C contains two excitations with very close energies, one is from $\Ket{d^{5}, J_{\text{eff}}=1/2}$ to $\Ket{d^{5}, J_{\text{eff}}=3/2}$ states and another is from $\Ket{d^{4},J_{\text{eff}}=0}$ to $\Ket{d^{4}, J_{\text{eff}}=2}$ states. Peak E is determined by the excitations from $\Ket{d^{4}, J_{\text{eff}}=0}$ to $\Ket{d^{4},J_{\text{eff}}=2^{\prime}}$. Obviously, our experimental observations have far richer features than this simulated \gls{RIXS} spectrum at isolated atom level. The activation of the inter-site hopping will strongly mix the $\Ket{d^4;d^5}$ and $\Ket{d^5;d^4}$ configurations, which will not only tune the energy of the excitations but also create new de-localized states and excitation channels. Here, we use $\Ket{d^{n_1};d^{n_2}}$ to represent a \gls{DPS} of the dimer, where Ir1 has $n_1$ electrons and Ir2 has $n_2$ electrons.

To estimate the hopping and the non-cubic crystal field we performed a first-principles \gls{DFT} calculation using \gls{VASP} \cite{kresse:1996,blochl:1994,kresse:1999,perdew:1996} and fit a $t_{2g}$ \gls{TB} Hamiltonian to the result using the maximally localized Wannier functions method~\cite{marzari:2012,mostofi:2008}.  We set $\lambda = U = J_H = 0$, as they are included explicitly in our model later. For simplicity we assume a trigonal local \gls{CF} for the IrO$_6$ octahedra in \BAIO{}. Under this approximation, the Hamiltonian $\hat{V}^{12}$ for the hopping between Ir1 and Ir2 takes a simple form in the $t_{2g}$ basis,
\begin{equation}
\hat{V}^{12}=\bordermatrix{
                           &d_{z_2x_2} &d_{z_2y_2} &d_{x_2y_2}\cr
                d_{z_1x_1} & -t        & t^{\prime}&-t\cr
                d_{z_1y_1} & t^{\prime}& -t        &-t\cr
                d_{x_1y_1} & -t        & -t        &t^{\prime}
} + h.c.
\end{equation}
where, $t$ and $t^{\prime}$ are the hopping parameters. Our \gls{DFT} calculation gives $t = 0.18$~eV, and $t^{\prime}\approx0.2t$. The on-site trigonal \gls{CF} Hamiltonian is
\begin{equation}
\hat{H}^{1(2)}_{\text{CF}}=
\begin{pmatrix}
\mu_{1(2)} & -\delta & -\delta\\
-\delta & \mu_{1(2)} & -\delta\\
-\delta & -\delta & \mu_{1(2)}\\
\end{pmatrix},
\end{equation}
where, $\mu_{1}$ and $\mu_{2}$ are the chemical potentials for Ir1 and Ir2, respectively. Their difference $\Delta \mu=\mu_{1}-\mu_{2}$ is about 0.1~eV. This chemical potential difference distinguishes the two inequivalent IrO$_6$ octahedras~\cite{Terzic2015coexisting,Streltsov_prb_2017_DFT_BAIO}, and induces partial charge disproportionation in \BAIO{}. 
$\delta$ is estimated to be 0.03 eV from the DFT calculation.
The two-site Ir1-Ir2 cluster Hamiltonian is  diagonalized in the subspace with 9 electrons in total to get the energy spectrum.

For the isolated Ir1 and Ir2 condition discussed earlier, the energy levels (0-1.2~eV) of these \gls{DPS} at $\Delta\mu=0$, $\delta=0$ and $t=0$ are shown as the red plateaus and labeled by S in Fig.~\ref{pic:theory}(a). After turning on the hopping $t$, dimer orbitals will form by superposition of the atomic orbitals of Ir1 and Ir2. As a result, the configurations $\Ket{d^4;d^5}$ and $\Ket{d^5;d^4}$ will mix with each other. With the \gls{DFT} derived $\{t, t^{\prime},\Delta\mu,\delta\}$, and the local interaction $J_{H}=0.3$~eV and $\lambda=0.345$~eV, the calculated energy levels of these delocalized dimer states are shown as blue lines in Fig.~\ref{pic:theory}(a).  As can be seen in Fig.~\ref{pic:theory}(c), the calculated \gls{RIXS} spectrum with these parameters agree quite well with our experimental observations. Peaks A and D, which were missing in the localized picture, now appear in the simulated \gls{RIXS} spectrum as excitations between delocalized dimerized states. For example, peak A is from the excitation from the ground state to the first excited state. The weights of $\Ket{d^4;d^5}$ and $\Ket{d^5;d^4}$ configurations are respectively 0.608 and 0.392 in the ground state and 0.560 and 0.440 in the first excited state.

To further appreciate the competition between the atomic \gls{SOC} $\lambda$, Hund's coupling $J_{H}$ and the delocalization hopping $t$, the evolution of the energy spectrum of these dimer states as a function of hopping $t$ is presented in Fig.~\ref{pic:theory}(b). In this calculation, other parameters except hopping $t$ are fixed. In Fig.~\ref{pic:theory}(b), the solid lines are the relative energies of the dimer states, and underlying colormap represents the calculated corresponding \gls{RIXS} intensities. We can see that each level S splits into many dimer states and the energy splitting increases as increasing $t$. In the small $t$ regime, local on-site $J_{H}$ and $\lambda$ dominate the Hamiltonian, so the mixing mainly occurs within the same level S and the energy splitting is not far away from the center of each level S. When increasing $t$, it will compete with $J_{H}$ and $\lambda$ to induce more mixing between different levels S, which are reflected as the crossing of energy levels in Fig.~\ref{pic:theory}(b). For example, the energy of the first excited state split from the level S0 increases at small $t$ regime to a maximum and then decreases at $t \gtrsim 0.12$. Another state split from level S1 is pushed down to cross with it and has very similar energy. 

Besides peaks A-E, the calculated \gls{RIXS} intensity also shows some shoulders around the main peaks which are not resolved in the experimental \gls{RIXS} spectrum. They might be washed out by longer range hopping or itineracy not captured in the cluster model. We also find that the energy of the dimer states and the \gls{RIXS} spectrum are not sensitive to $\Delta\mu$ in the region $\Delta\mu < t$ [see Fig.~\ref{pic:theory}(c)], because $t$ dominates the energy spectrum in this region and the energy splitting of the bonding and anti-bonding orbitals is as large as $2.5t$. At $\Delta\mu=0.1$~eV and $t=0.18$~eV, the calculated charge disproportionation between Ir1 and Ir2 is about 0.218 electrons, which is close to 0.3 electrons reported in previous DFT calculation~\cite{Streltsov_prb_2017_DFT_BAIO}.

Combining \gls{RIXS} measurements and a \gls{DFT} calculations, we show that the electronic structure of \BAIO{} needs to be described by partially-delocalized dimer orbitals, rather than $J_\text{eff}=1/2$ atomic states in the strong \gls{SOC} limit. It is interesting to notice that dimer orbtials can still occur in \BAIO{} even if $U$ is large, as due to the non-integer average Ir valence of 4.5, the Mott mechanism cannot stop hopping between the Ir sites. The dimerization process significantly changes the magnetic properties of \BAIO{}. The calculated effective local moment per dimer is about 1~$\mathrm{\mu_{B}}$ in \BAIO{}, which is consistent with the experimental results~\cite{Terzic2015coexisting}, but is much smaller than the value 1.732~$\mathrm{\mu_{B}}$ expected for two isolated Ir sites: a $J_{\text{eff}}=1/2$ on Ir2 and a $J_{\text{eff}}=0$ singlet on Ir1. It indicates that the symmetry of the magnetic order parameter has deviated from the ideal spherical symmetry. Indeed, unlike $3d$ magnetic Mott insulators, where the order parameter can be described by pure local physics, it is common that the order parameter cannot be well described by a pure local object in many iridates due to the extended orbitals and much stronger inter-site hoppings. 
For instance, reduced effective local moments have been observed in many $d^{5}$ iridates, such as Sr$_2$IrO$_4$~\cite{Gao_prb_1998,Kimprl2008}, Sr$_3$Ir$_2$O$_7
$~\cite{Cao_prb_2002} and pyrochlore iridates~\cite{Ishikawa2012,shapiro2012,disseler:2012}. Another interesting case is the exotic magnetic moments found in some $d^{4}$ iridates~\cite{cao_prl_2014} where the local ground state is a non-magnetic singlet. This peculiar behavior is found to be caused by the virtual inter-site exchange process~\cite{Khaliullin2013}, so a local single atomic description is not appropriate here either. We also emphasize that the cluster model used in this work can be straightforwardly applied to simulate \gls{RIXS} spectrum of the 6H-hexagonal oxides Ba$_3AB_2$O$_9$ ($A$=In, Y, Lu, Na and $B$=Ru, Ir)~\cite{Ba3InIr2O9_2017,Ba3MRu2O9_2017,Ba3NaRu2O9_2012} and similar dimer excitations can be expected in these compounds. This theoretical method can also be generalized to study other systems with strong nonlocal electronic itinerancy, such as the pyrochlore iridates, where possible low energy dimer excitations (around 0.1~eV) have been observed in \gls{RIXS} spectrum \cite{Clancy_2016,donnerer_2016,chun_2018,Calder2016}.

In summary, we find new peaks corresponding to the excitations of dimer orbitals in the experimental \gls{RIXS} spectrum of \BAIO{}. The \gls{DFT} calculations and the \gls{RIXS} simulations confirm that the hopping strength between Ir1 and Ir2 in \BAIO{} is indeed strong enough to form dimer orbitals and their excitations can be seen in the \gls{RIXS} spectrum. We demonstrate that this combination of experimental and theoretical \gls{RIXS} study can directly confirm the existence of delocalized dimer orbitals in \BAIO{} and be used to explain the observed reduction of magnetic moment. This serves as an example and may open new directions to study  delocalization in other dimerized strongly correlated materials by \gls{RIXS} with widespread potential applications in molecules, oxide heterostructures and even ultra-fast transient states.

\begin{acknowledgments}
We thank John Hill, Sergey Streltsov, Hu Miao and Keith Gilmore for discussions and support related to this project. Work at ShanghaiTech U. was partially supported
by MOST of China under the grand No. 2016YFA0401000. R. W. was supported by international partnership program of Chinese Academy of Sciences under the grand No.112111KYSB20170059. This work received finical support from the U.S.\ Department of Energy, Office of Basic Energy Sciences, Early Career Award Program under Award Number 1047478. Work at Brookhaven National Laboratory was supported by the U.S.\ Department of Energy, Office of Science, Office of Basic Energy Sciences, under Contract No. DE-SC0012704. Work at Argonne is supported by the U.S. Department of Energy, Office of Science, under contract No.\ DE-AC-02-06CH11357. Y.W. and G.K.\ was supported by the US Department of energy, Office of Science, Basic Energy Sciences as a part of the Computational Materials Science Program through the Center for Computational Design of Functional Strongly Correlated Materials and Theoretical Spectroscopy. G.C.\ acknowledges the NSF support via NSF grant DMR-1712101. 
\end{acknowledgments}

\bibliography{dimer_rixs}

\end{document}